\documentclass[aps,superscriptaddress,12pt]{revtex4}

\usepackage{times}
\usepackage{graphicx}
\usepackage{moreverb}
\usepackage{amssymb,amsmath}

\def\d{\mathrm{d}}
\def\i{\mathrm{i}}
\def\e{\mathrm{e}}

\begin{document}

\title{
\ \\
Classical and Quantum Fisher Information in the Geometrical Formulation of Quantum Mechanics
}

\author{Paolo Facchi} 
\affiliation{Dipartimento di Matematica, Universit\`a di Bari, I-70125 Bari, Italy}
\affiliation{INFN, Sezione di Bari, I-70126 Bari, Italy}
\affiliation{MECENAS, Universit\`a Federico II di Napoli \& Universit\`a di Bari, Italy}

\author{Ravi Kulkarni} 
\affiliation{Vivekananda Yoga Research Foundation, Bangalore 560 080, India}

\author{V. I. Man'ko} 
\affiliation{P. N. Lebedev Physical Institute, Leninskii Prospect 53, Moscow 119991, Russia}

\author{Giuseppe Marmo} 
\affiliation{Dipartimento di Scienze Fisiche, Universit\`a di Napoli ``Federico II'', I-80126 Napoli, Italy} \affiliation{INFN, Sezione di Napoli, I-80126 Napoli, Italy}
\affiliation{MECENAS, Universit\`a Federico II di Napoli \& Universit\`a di Bari, Italy}

\author{E. C. G. Sudarshan} 
\affiliation{Department of Physics, University of Texas, Austin, Texas 78712, USA}

\author{Franco Ventriglia} 
\affiliation{Dipartimento di Scienze Fisiche, Universit\`a di Napoli ``Federico II'', I-80126 Napoli, Italy}
\affiliation{INFN, Sezione di Napoli, I-80126 Napoli, Italy}
\affiliation{MECENAS, Universit\`a Federico II di Napoli \& Universit\`a di Bari, Italy}

\date{\today}

\begin{abstract}

The tomographic picture of quantum mechanics has brought the description of quantum states closer to that of classical probability and statistics. On the other hand, the geometrical formulation of quantum mechanics introduces a metric tensor and a symplectic tensor (Hermitian tensor) on the space of pure states. By putting these two aspects together, we show that the Fisher information metric, both classical and quantum, can be described by means of the Hermitian tensor on the manifold of pure states.
\end{abstract}

\maketitle

\pagebreak

\section{Introduction}

The states of quantum systems are described by wave functions (state
vectors in Hilbert space) or density matrices. The difference between quantum
states can be associated with a distance between the state vectors or the
density matrices.
To introduce the notion of the distance one needs to construct a metric
in the set of states.
In classical probability theory the Fisher information
metric can be used to characterize the distance between probability
distributions.
In quantum information theory the quantum generalization of the metric is
also used.

In the past two decades, the tomographic picture of Quantum Mechanics has shown that quantum states may be described by means of genuine probability distributions, called tomograms \cite{tomog}. This allows the use of methods of classical probability theory to deal with quantum states. Of course, the converse is also possible and we can view classical probability theory within the quantum setting. We shall consider this second possibility to express the Fisher classical information metric within the quantum paradigm. In doing this we obtain that the appropriate expression contains the Quantum Information Metric and reduces to the classical one when states satisfy suitable conditions.

More specifically, in classical optics, photometry dominates the measured quantities. In
radiative transfer we must include the direction cosines of light rays
as well as the spectrum. But even the two slit interference demands a
phase (or rather phase differences). This is also true for the
description of partial coherence. Pancharatnam showed that propagation
in crystals also requires the introduction of a phase for the
wavefunction. This notion was amplified  by Berry by introducing a
path dependent phase. (Already Dirac, when dealing with the magnetic monopole,had introduced phase dependence on the path.) In all these cases the primary measurement is of
intensities only and he showed that such a phase is present in general
in quantum mechanics.
So classical intensity distribution is insufficient for a complete
description. Given a classical (non-negative, normalised) probability
we should introduce a phase.

The main observation is the following: we describe probability densities $p(x)$ of random variables with values in $X$ by means of probability amplitudes, i.e. normalized wave functions $\psi(x)$ defined on $X$, by setting $p(x) = \psi^{*}(x)\psi(x) = |\psi(x)|^2$, thus going from integrable functions to square integrable functions on $X$ \cite{BH}.
Our strategy consists of using the available metric tensors on $\mathcal{H}$ and thereof on the space of pure states $\mathcal{R}(\mathcal{H})$ and to pull them back to a submanifold $\Theta$ of probability densities over $X$.

We shall find that the Hermitian tensor fields on $\mathcal{R}(\mathcal{H})$ when pulled back to $\Theta$ will give the Fisher Quantum Information metric tensor. The aim of this work is to exhibit explicitly the form of this metric tensor on the space $\Theta$ starting with the Fubini-Study metric on the space of pure states.

\section{The metric tensor on the space of pure states}

It is well known that due to the probabilistic interpretation, states for quantum systems are \em not \em vectors $| \psi \rangle \in \mathcal{H}$ but rather they are rays, elements of the Hilbert manifold $\mathcal{R}(\mathcal{H})$, which are conveniently parametrized as rank-one projection operators, the projection from $\mathcal{H}$ to $\mathcal{R}(\mathcal{H})$ being defined by 
\begin{equation}
\pi: |\psi\rangle \mapsto \frac{| \psi \rangle \langle \psi |}{\langle \psi | \psi \rangle},
\end{equation}
for $| \psi \rangle\neq 0$.
This projection map allows to identify on $\mathcal{R}(\mathcal{H})$ a metric tensor $\mathfrak{g}$ usually called the Fubini-Study metric and a symplectic structure $\omega$ \cite{reviewBNGJ}. Both of them define on $\mathcal{R}(\mathcal{H})$ what is called a K\"ahlerian structure. The pullback of this tensor to $\mathcal{H}$ along the map $\pi$ acquires the following form:
\begin{equation}
\label{eq:FSmetric}
\mathfrak{h} = \frac{\langle \d\psi 
| \d\psi \rangle}{\langle \psi | \psi \rangle} - \frac{\langle \d\psi | \psi \rangle\, 
\langle \psi | \d\psi \rangle}{\langle \psi | \psi \rangle^2}
\end{equation}
as has been shown elsewhere \cite{entanglebook}. To work with this tensor on $\mathcal{H}$ instead of $\mathcal{R}(\mathcal{H})$ is quite convenient for computational purposes. 

Let the Hilbert space $\mathcal{H}$ be realized as the space of square integrable functions over $X$, namely $\mathcal{H}=L^2(X)$. Therefore, abstract vectors $|\psi\rangle$ are wave functions $\psi(x)$, and their scalar product is $\langle \psi | \phi \rangle=\int_X \psi(x)^* \phi(x)\, dx$.

The physical state $|\psi\rangle$ may depend on unknown parameters $\theta_1, \theta_2,..., \theta_m$ and this can be made explicit using the notation $\psi(x;\theta)$ for the wave function. This will be the setting for what follows. 
Having replaced $|\psi\rangle$ with a wave-function $\psi(x;\theta)$ we can consider a polar representation by setting 
\begin{equation}
\label{eq:polarrep}
\psi(x;\theta) = p(x;\theta)^{1/2} \e^{\i\alpha(x;\theta)},
\end{equation}
with $\langle \psi | \psi \rangle = 1$, so that $p\in L^1(X)$ is a probability density.

We should say something about our notation. The probability density $p(x;\theta)$ is being used to consider averages of functions
\begin{equation}
\mathbb{E}_p(f) := \int_X{f\, p dx} ,
\end{equation}
averages of differential forms 
\begin{equation}
\mathbb{E}_p(\d f) := \int_X{\d f \,pdx},
\end{equation} 
and more generally averages of covariant tensors like $\int{\d f\, \d g \,pdx }$. 

For instance, if $f$ depends on parameters $(\theta_1, \theta_2, ..., \theta_m)$ we think of $\d f$ as 
\begin{equation}
\d f = \sum_{k=1}^{m}{\frac{\partial f}{\partial \theta_k}d\theta_k}
\end{equation}
and similarly
\begin{equation}
\d f \, 
\d g = \sum_{k=1}^{m}\sum_{j=1}^{m}{\frac{\partial f}{\partial \theta_j}\frac{\partial g}{\partial \theta_k} \d\theta_j \, 
\d\theta_k} .
\end{equation}
The advantage of using the abstract notation $\d f$ is that we do not have to specify the parameters or their number. Moreover, from the abstract notation, we would have $|\d\psi\rangle$ and $\langle x | \d \psi \rangle = \d\psi(x)$, showing that the differential should \emph{not} be understood as taken with respect to ${x}$ which identifies an orthonormal basis of improper eigenvectors which are considered to be chosen once and for all.

Using the polar representation (\ref{eq:polarrep}) above for $\psi(x)$ we have $\d(\ln\psi(x;\theta)) = \frac{1}{2}\d(\ln p(x;\theta)) + \i\,\d\alpha(x;\theta)$, while the normalization condition implies that $\langle \d\psi | \psi \rangle = -\langle \psi | \d\psi \rangle$. 

Using expression (\ref{eq:FSmetric}) for $\mathfrak{h}$, we obtain for the pullback of $\mathfrak{h}$, denoted by $\mathfrak{h}_X$, the expression
\begin{eqnarray}
\mathfrak{h}_X &=& \frac{1}{4}\int_X{(\d\ln p)^2 \,pdx} + \int_X{(\d\alpha)^2\,pdx} - 
\left(\int_X{\d\alpha\,pdx}\right)^2 \nonumber\\
& & - \i\int_X{(\d\ln p\; \d\alpha - \d\alpha \;  \d\ln p)\,pdx} .
 \label{eq:pullbackh}
\end{eqnarray}
We have used a few identities in deriving this expression which follow from $\int{pdx} = 1$, namely $\int{\d p \, dx } = \int{\d\ln p \, p dx } = 0$.

From (\ref{eq:pullbackh}) we obtain for the metric tensor $\mathfrak{h}_X$ the expression
\begin{equation}
\mathfrak{h}_X = \mathfrak{g}-\i\,\omega,
\end{equation}
where
\begin{equation}
\mathfrak{g}=\frac{1}{4}\mathbb{E}_p\left[(\d \ln p)^2\right] + \mathbb{E}_p\left[(\d\alpha)^2\right] - 
\left[\mathbb{E}_p(\d\alpha)\right]^2, \qquad
\omega= \mathbb{E}_p\left[\d \ln p \wedge \d\alpha \right].
\end{equation}

This Hermitian tensor on $\Theta$ coincides with the Fisher classical information metric when $\d\alpha = 0$. To see this,  consider a parameter space $\Theta \equiv \{\theta_1, \theta_2, ..., \theta_m\}$. If we compute our metric tensor $\mathfrak{h}_X$ on contravariant vectors $\frac{\partial}{\partial\theta_j}$, $\frac{\partial}{\partial\theta_k}$ we obtain
\begin{eqnarray}
(\mathfrak{h}_X)_{jk} &=& \mathfrak{h}_X\left(\frac{\partial}{\partial\theta_j},\frac{\partial}{\partial\theta_k}\right)
\nonumber\\
&=& \frac{1}{4}\mathcal{F}_{jk} + \mathbb{E}_p\left(\frac{\partial \alpha}{\partial \theta_j}\frac{\partial \alpha}{\partial \theta_k}\right) - 
\mathbb{E}_p\left(\frac{\partial \alpha}{\partial \theta_j}\right)\mathbb{E}_p\left(\frac{\partial \alpha}{\partial \theta_k}\right) - \i\,\mathbb{E}_p\left(\frac{\partial \ln p}{\partial \theta_j}\frac{\partial \alpha}{\partial \theta_k} - \frac{\partial \ln p}{\partial \theta_k}\frac{\partial \alpha} {\partial \theta_j}\right), \quad
\label{eq:h=F+alpha}
\end{eqnarray}
where
\begin{equation}
\mathcal{F}_{jk} = \mathbb{E}_p\left(\frac{\partial \ln p}{\partial \theta_j} \frac{\partial \ln p}{\partial \theta_k}\right) 
\end{equation}
is the Fisher classical information metric, whose abstract expression reads
\begin{equation}
\mathcal{F} = \mathbb{E}_p\left[(\d \ln p)^2\right].
\end{equation}

It is clear that the second and the third terms in (\ref{eq:h=F+alpha}) combine to give the covariance of $\d \alpha$ and that the imaginary part of (\ref{eq:h=F+alpha}) is connected with the geometric phase. So when $\mathrm{Cov}(\d \alpha)$ and the geometric phase are both zero,
we recover the Fisher classical information metric, namely
\begin{equation}
\mathfrak{h}_X =\frac{1}{4} \mathcal{F}.
\end{equation}
In general, we have that the Fisher classical information metric $\mathcal{F}/4$ is strictly dominated by the quantum Riemannian metric $\mathfrak{g}$ \cite{reviewBNGJ}.

In the general case ($\d\alpha \neq 0$) $\mathfrak{h}_X$ coincides with the Fisher quantum information metric. This will be shown in the next section.

\section{Fisher quantum information metric}
A definition of the Fisher quantum information metric was proposed by Helstrom \cite{helstrom}. This definition relies on the notion of the symmetric logarithmic derivative. The symmetric logarithmic differential $\d L_{\rho}$ is implicitly defined by the relation
\begin{equation}
\d\rho = \frac{1}{2}\left(\rho\, \d L_{\rho} + \d L_{\rho}\,\rho\right),
\label{eq:SLD}
\end{equation}
where $\rho$ represents a generic density matrix (which we prefer to call a density state) and $\d L_{\rho} = \d L_{\rho}^{\dagger}$ defines the Hermitian matrix whose matrix elements are differential one-forms. The uniqueness of $\d L_{\rho}$ may be proved by adopting the arguments in \cite{holevobook}, p.~274. The Fisher quantum information acquires the form 
\begin{equation}
\mathcal{F}_q = \mathrm{Tr} \left[\rho\, (\d L_{\rho})^2 \right].
\end{equation}
As usual the trace replaces the integrals which appear when we consider probability distributions.

By restricting our computations to pure states, i.e. $\rho^2 = \rho$, $\rho^\dagger = \rho$, $\mathrm{Tr} \rho = 1$, we find the identities 
\begin{equation}
i)\, \rho\, \d\rho + \d\rho\, \rho = \d\rho, \qquad
ii)\, \mathrm{Tr} (\d\rho) = 0, \qquad
iii)\, \mathrm{Tr} (\rho\, \d\rho) = 0 .
\end{equation}
From the definition of the symmetric logarithmic differential (\ref{eq:SLD}) compared with i) we find that
\begin{equation}
\d L_{\rho} = 2\d\rho.
\end{equation}
Thus for pure states we get
\begin{equation}
\mathcal{F}_q = 4 \mathrm{Tr} \left[\rho\, (\d \rho)^2\right].
\end{equation}
We recall that by the differential of a matrix we mean a matrix-valued differential one-form, i.e. the matrix which we obtain by taking the differentials of the elements of the matrix.

To carry out the comparison of $\mathcal{F}_q$ with $\mathfrak{h}_X$, we start with
\begin{equation}
\rho = |\psi\rangle\langle\psi|, \qquad
\langle\psi|\psi\rangle = 1, \qquad
\langle \d\psi|\psi\rangle = - \langle\psi|\d\psi\rangle
\end{equation}
From $\d\rho = |\d\psi\rangle\langle\psi| + |\psi\rangle\langle \d\psi|$ we compute easily 
\begin{equation}
\mathrm{Tr} \left[\rho\, (\d\rho)^2\right] = \langle \d\psi|\d\psi\rangle - \langle \d\psi|\psi\rangle \, \langle\psi|\d\psi\rangle
\end{equation}
which is exactly our tensor field $\mathfrak{h}$, given in (\ref{eq:FSmetric}) when $\langle\psi|\psi\rangle = 1$.

In conclusion, we have found that for pure states, what we have called the Fisher quantum information metric contains both the quantum version and the classical version when $\d\alpha = 0$.

\section{Conclusions and outlook}

Much interest has been focused on the quantum counterpart of the classical Fisher information \cite{braunstein-caves}. The quantum counterpart of the classical Fisher information was shown to constitute an upper bound on the classical Fisher information. Consequently there was interest in understanding conditions under which the bound could be attained. Barndorff-Nielsen and Gill \cite{bngill} derived a condition for the quantum and classical Fisher information to coincide in the particular case of a two-dimensional pure state system. Luati \cite{luati} showed that this condition held even for two-dimensional mixed states. Our geometrical formulation of the quantum Fisher information shows that the condition for the equality of the quantum and classical information is the condition $d\alpha = 0$ for pure states in any dimension.

We will elsewhere discuss the implications of our geometrical formulation of Fisher information in terms of the Fubini-Study metric and tomographic probabilities.

Our presentation of Fisher quantum information metric is closer in spirit to
what is known in the literature as ``nonparametric'' Fisher information
metric \cite{streater}. In our approach however we consider a manifold of states suitably chosen so that it carries a differential calculus.

An additional merit of our description is that we consider probability
amplitudes instead of probability densities, therefore it is possible to work
on $\mathcal{H}$ rather than on $\mathcal{R(H)}$, this means we can deal with $L^{2}$-spaces
instead of $L^{1}$-spaces. These considerations will be quite useful later on when we
move from pure states to generic density states. In our approach,the
classical Fisher information metric is recovered by restricting the
imbedding into a Lagrangian subspace of $\mathcal{H}$. In a future paper we shall
consider the available geometric picture of the Gelfand-Naimark-Segal
construction \cite{GNS} to extend our approach to the $C^*$-algebraic
 approach for statistical models elaborated by Streater, and to compare
more closely our approach to the one by Gibilisco and Isola \cite{Gibilisco}.

We believe that our present treatment will be relevant to further enhance
geometrical methods in the analysis of statistical models, both from the
conceptual point of view and the methodological point of view as well.

\end{document}